\def\ube13{UBe$\rm_{13}$}

\def\bi2212{Bi$\rm_2$Sr$\rm_2$CaCu$\rm_2$O$\rm_8$}
\def\ybi2212{Bi$\rm_2$Sr$\rm_2$YCu$\rm_2$O$\rm_8$}
\def\ycabi2212{Bi$\rm_2$Sr$\rm_2$Ca$\rm_{1-x}$Y$\rm_x$Cu$\rm_2$O$\rm_{8+\delta}$}

\def\y65cabi2212{Bi$\rm_2$Sr$\rm_2$Ca$\rm_{0.35}$Y$\rm_{0.65}$Cu$\rm_2$O$\rm_{8+\delta}$}

\def\Co{CeCoIn$_5$}

\documentstyle[aps,prb,twocolumn,epsf]{revtex}
\begin{document} 
\draft

\def\dfrac#1#2{{\displaystyle{#1\over#2}}}
\twocolumn[\hsize\textwidth\columnwidth\hsize\csname @twocolumnfalse\endcsname
Submitted to Phys. Rev. Lett.\hfill{LA-UR-016718}

\title{First order superconducting phase transition in CeCoIn$_5$.}

\author{A. Bianchi,$^1$ R. Movshovich,$^1$  N. Oeschler,$^2$ P. Gegenwart,$^2$ F. Steglich,$^2$ J. D. Thompson,$^1$ P. G. Pagliuso,$^1$ and J.~L.~Sarrao$^1$ }
\address{$^1$Los Alamos National Laboratory, Los Alamos, New Mexico 87545 \\ $^2$ Max-Planck-Institute for the Chemical Physics of Solids, Noethnitzer Str. 40, 01187 Dresden, Germany}

\date{\today}

\maketitle

\begin{abstract} 

We investigated the magnetic field dependence of the superconducting phase transition in heavy fermion \Co\ (T$_{c} = 2.3$ K) using specific heat, magneto-caloric effect, and thermal expansion measurements. The superconducting transition becomes first order when the magnetic field is oriented along the 001 crystallographic direction with a strength greater that 4.7 T, and transition temperature below $T_0 \approx 0.31 T_c$.  The change from second order at lower fields is reflected in strong sharpening of both specific heat and thermal expansion anomalies associated with the phase transition, a strong magnetocaloric effect, and a step-like change in the sample volume. The first order superconducting phase transition in \Co\ is caused by Pauli limiting in type-II superconductors, and was predicted theoretically in the mid 1960s. We do not see evidence for the inhomogeneous Fulde-Ferrell-Larkin-Ovchinnikov (FFLO) superconducting state (predicted by an alternative theory also dating back to mid-60's) in \Co\ with field $H \parallel$ 001.


\end{abstract}


]
\narrowtext

The behavior of superconductors in magnetic field has primary scientific and technological importance. It underlies such diverse areas as magnetic imaging, energy transmission and storage, ultrasensitive instrumentation and electronics, and many other fields of technology and medicine. At the same time it reflects a very fundamental property of matter - the behavior of electrons in magnetic field. BCS theory, presented in 1957 (Ref.~\cite{bcs}), gave microscopic explanation of a number of phenomena observed during the previous half century of research on superconductivity. The theories put forth in the early 1960's, that addressed the effect of magnetic field on superconductivity, were the first extensions of BCS that made predictions of new phenomena and provided tests of BCS theory's predictive powers~\cite{maki:ptp_64,maki:pr_66,fulde_ferrell:pr_64,larkin_ovchinnikov:jetp_64,gruenberg:prl_66}. Magnetic field can suppress superconductivity via two effects: orbital pair breaking of superconducting pairs in the superconducting state and Pauli paramagnetism due to electron spins, which lowers the relative energy of the normal state. It was shown that when the Pauli effect is sufficiently strong relative to the orbital effect, the superconducting phase transition may change from second order (BCS result for zero field) to first order\cite{maki:ptp_64,maki:pr_66}. This is due to competition between two energies basic to the understanding of superconducting and normal states of metals: condensation energy of superconducting pairs and magnetic energy of the normal electron spins due to Pauli paramagnetism. This prediction is very straightforward, and yet eluded confirmation for almost 40 years. This is partly the reason why the experimental observation of the first order superconducting transition became one of the ``holy grails" of the superconducting community. 

A number of conventional superconductors were proposed as candidates for observation of the first order superconducting transition in magnetic field, due to their high orbital critical field $H_{c20}$ (weak orbital pair breaking) and, therefore, relatively strong Pauli limiting effect, in the early and mid-sixties. Experimental search, however, came up with null results~\cite{berlincourt:pr_63,kim:pr_65,shapira:pr_65,hake:prl_65}. The failure to observe the first order superconducting transition was attributed to high spin-orbit scattering rate in all of the compounds investigated~\cite{maki:pr_66}. Here we present specific heat, magnetocaloric, thermal expansion, and magnetostriction data for \Co\ that demonstrate that the superconducting phase transition indeed changes from second to first order at a critical point with temperature $T_0 = 0.31 T_c$, in very good agreement with theoretical estimate. Recently, a first order phase transition at low temperature in \Co\ was inferred from the step in thermal conductivity in \Co\ at $H_{c2} \parallel [001]$, which was suggested to be ``likely due to an entropy jump"~\cite{izawa:prl_01}. Our specific heat data offer a direct and unambiguous proof of the first order phase transition in \Co\ with magnetic field in this orientation.

\Co\ is a recently discovered ambient pressure heavy fermion superconductor~\cite{petrovic:jpcm_01} with the record high superconducting transition temperature $T_c = 2.3$ K for this class of compounds. A number of thermodynamic and spectroscopic measurements indicate spin-singlet, even pairing state, with lines of nodes in the superconducting energy gap~\cite{movshovich:prl_01,kohori:prb_01,movshovich:sces_01}. Therefore, the lowest possible symmetry of the superconducting state of \Co\ is d-wave. Recently, Izawa et al.~\cite{izawa:prl_01} reported the four-fold modulation of thermal conductivity of \Co\ in magnetic field in support of the $d_{x^2-y^2}$ order parameter, similar to high temperature superconductors.

First order superconducting phase transition in \Co\ occurs in magnetic field close to the superconducting critical field $H_{c2} = H_{c2}(T = 0) = 4.95$ T, with field along the [001] crystallographic direction. The 

\begin{figure}
\epsfxsize=3in
\centerline{\epsfbox{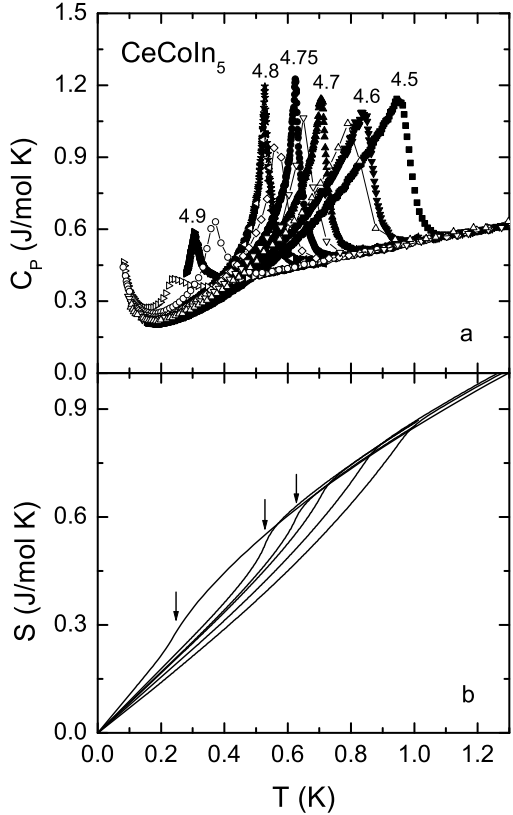}}
\caption{
Specific heat and entropy of \Co. (a) Specific heat vs. temperature. Closed symbols - decay method, numbers over the peaks indicate magnetic field in T. Open symbols - heat the pulse method. ($\bigtriangleup$) 4.62 T, ($\bigtriangledown$) 4.72 T, ($\diamond$) 4.77 T, ($\triangleleft$) 4.8 T, ($\circ$) 4.87 T, ($\triangleright$) 4.925 T. (b) Entropy as a function of temperature from (a). From left to right: 4.925 T, 4.8 T, 4.75 T, 4.7 T, 4.6 T, 4.5~T. Arrows indicate step-like features in entropy at $T_c$ for $H > 4.7$ T.}
\label{hc_T_sweep}
\end{figure}

\noindent
change from the second order nature of the transition, observed at zero and low magnetic field, to first order at high field occurs at $T_0 \approx 0.72{\rm\ K} = 0.31 T_c$. We used a variety of techniques to investigate the region of interest of the $H - T$ phase diagram, including measurements of specific heat at constant magnetic field, magnetocaloric effect, thermal expansion, magnetostriction, and resistivity. 

Fig.~\ref{hc_T_sweep}(a) shows specific heat as a function of temperature for several values of magnetic field in the high 
field region. The data represented with open symbols were collected with the standard heat pulse technique. When the phase transition becomes first order, it sharpens substantially. In this regime the temperature decay method, where specific heat is extracted directly from the temperature trace of the system coming to equilibrium, was particularly useful to resolve the specific heat anomaly (solid symbols). The two data sets for $H = 4.8$ T, obtained with the temperature decaying down (stars) and up (dots), as well as the data collected with heat pulse method (left open triangles) overlap each other, indicating good internal thermal equilibrium of the cell. 

The specific heat anomaly for the second order phase transition (e.g. 4.5 T data) displays a characteristic step at $T_c$ (predicted to be $\Delta C / C = 1.43$ in the BCS theory), and then gradually drops below $T_c$. In \Co\ $\Delta C / C$ drops steadily from the very high value of 4.5 at zero field~\cite{petrovic:jpcm_01} to $\approx 1.1$ at $H = 4.5 \, \rm T$. The temperature width of the anomaly at half maximum $\Delta T_{HM}$ scaled by $T_c$ for 4.5 T is $w (4.5 \,\rm T) = \Delta T_{HM} / T_c \large|_{4.5 \,\rm T} \approx 25$\%.

As we move into the regime of the first order phase transition, the maximum of the specific heat anomaly rises, e.g. $\Delta C / C (4.8 \,\rm T) = 1.9$, and it becomes substantially narrower, $w (4.8 \,\rm T) = 6$\%. It should be kept in mind that this sharpening of the anomaly takes place in the region of the phase diagram where the boundary between the normal and superconducting states is crossed at more of a glancing angle during the temperature sweep as $H \rightarrow H_{c2}$. As a result, the anomaly should broaden if it remains second order, in contrast to the evolution of the data. 

Figure~\ref{hc_T_sweep}(b) shows the entropy $S$ of \Co\ as a function of temperature for various values of the magnetic field, obtained from the data in Fig.~\ref{hc_T_sweep}(a). There is a clear difference in the temperature evolution of the entropy for the field below and above $H_0 \approx 4.7$ T. Below this field ($T_c > 0.7$ K) transition manifest itself by a kink in $S$. For fields  above 4.7 T there is a step-like feature in $S$ at $T_c$, as expected for a first order phase transition. 

To elucidate the low temperature behavior and determine the temperature $T_0$ of the critical point (and corresponding field $H_0$) at which the superconducting transition changes from second to first order, with higher precision, we studied the phase diagram of \Co\ by sweeping magnetic field. This was done under close to adiabatic (constant entropy) conditions, regulating the bath temperature to be approximately that of the sample, which was weakly thermally coupled to the bath. The behavior of the system is then governed by the magnetocaloric effect.

\begin{figure}
\epsfxsize=3in
\centerline{\epsfbox{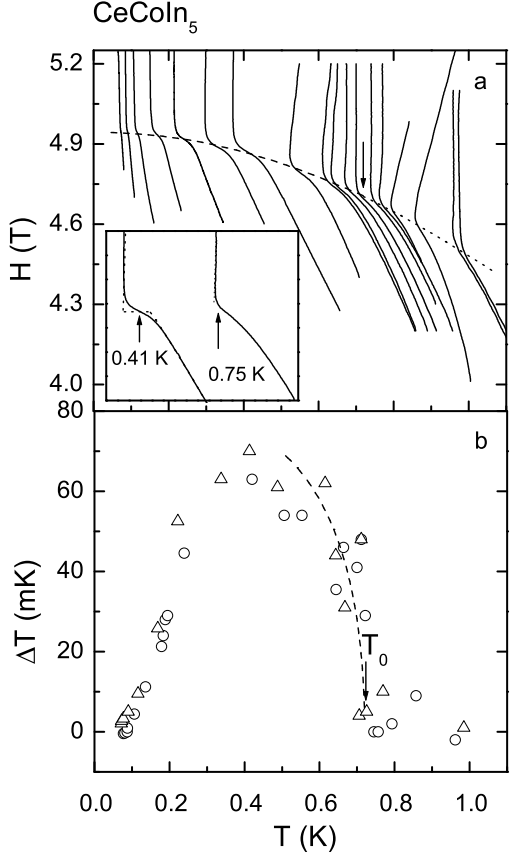}}
\caption{
Quasi-adiabatic sweeps of magnetic field. (a) H vs. T during magnetic field sweep for various starting (T, H) points. Dotted (dashed) line is a phase boundary in second (first) order transition region. Arrow - same as in (b). Inset: magnetocaloric field sweep with $\rm T_c = 0.41 {\rm \ K}< T_0$ and $\rm T_c = 0.74{\rm \ K} > T_0$. Dotted lines are cubic fits to the data and $\Delta T_{pb}$ for data with $T_c$ = 0.74 K. (b) Change in temperature at the phase transition during magnetic sweeps in (a). ($\circ$) field swept up, ($\bigtriangleup$) field swept down. Arrow indicates the temperature $T_0$ at which the superconducting transition changes from first to second order. Dotted line - guide to the eye in the region of the first order transition close to $T_0$.} 
\label{magnetocaloric}
\end{figure}

Figure~\ref{magnetocaloric}(a) shows several sweeps of magnetic field up and down, starting at different temperatures. Below $H_0$ there is a sharp change in temperature as magnetic field crosses the phase boundary. Temperature drops when the phase boundary is crossed as the field is swept up, since the system goes from the low entropy to the high entropy phase, and the temperature has to decrease to keep the entropy constant~\cite{jaime:nature_00}. The temperature swing is reversed (temperature rises) on the down field sweep. The change in the temperature of the sample after the crossing of the first order phase boundary $\Delta T_{pb}$ is a measure of the latent heat associated with transition. $T_c$ corresponds to the maximum of the derivative $dT \over dH$, and $\Delta T_{pb}$ is determined by extrapolations of the fits to the $H$ vs. $T$ curves outside of the transition region to $T_c$. An example of such a procedure is displayed in the inset of Fig.~\ref{magnetocaloric}(a) for the data with $T_c = 0.41$ K, where horizontal dotted line segment represents $ \Delta T_{pb}$. The difference $\Delta T_{pb}$ should equal zero 

\noindent
for the second order phase transition. This is indeed observed, within experimental scatter, for the data with $T_c > T_0$, illustrated in the inset of Fig.~\ref{magnetocaloric}(a) for the data with $T_c$ = 0.74 K. Figure~\ref{magnetocaloric}(b) displays the measured $\Delta T_{pb}$ as a function of $T_c$. The cross-over from the first order transition with non-zero $\Delta T_{pb}$ to the second order transition with $\Delta T_{pb} = 0$ occurs at a sharply defined critical temperature $T_0 = 0.72 \pm 0.02$ K, which is indicated by the arrow.

\begin{figure}
\epsfxsize=3in
\centerline{\epsfbox{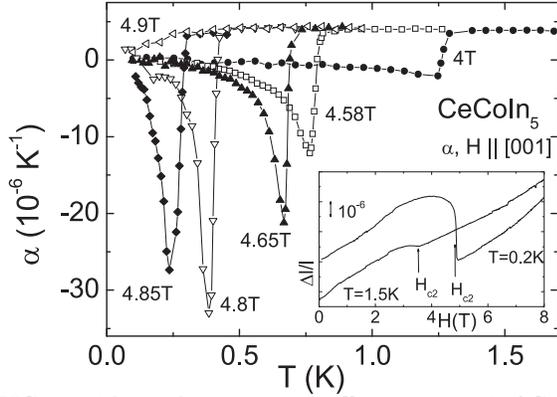}}
\caption{
Thermal expansion coefficient $\alpha$ vs. $T$ of \Co\ for fields $4\,\mbox{T} \leq H \leq
4.9\,\mbox{T}$ applied along [001]. The inset shows the relative length change $\Delta l/l$ vs. $H$ at $T=1.5\,\mbox{K}$ and $T= 0.2\,\mbox{K}$. The displayed curves represent raw data without background subtraction. The arrows indicate anomalies related to $H_{c2}$.}
\label{thermal expansion}
\end{figure}

We also observed the change from the second to first order nature of the superconducting transition of \Co\ via thermal expansion measurements, depicted in Fig~\ref{thermal expansion}. The coefficient of the thermal expansion, $\alpha(T) = l^{-1} dl/dT$, was determined down to 50 mK by utilizing an ultrahigh-resolution capacitive dilatometer with a maximum sensitivity of $\Delta l/l \geq 10^{-11}$. Thermal expansion along the crystallographic [001] direction was measured for two different plate-like single crystals in magnetic fields up to 8 T applied along [001]. For one of the crystals isothermal magnetostriction measurements were performed as well at $T=0.2\,\mbox{K} $ and at $T=1.5\, {\mbox K}$.

In the low field range $\rm H \leq 4\,$ T a step-like anomaly in $\alpha$, indicative of a second-order
transition, is observed, which shifts towards lower temperatures upon increasing $H$. With increasing fields (see Fig.~\ref{thermal expansion}) the signature in $\alpha$ sharpens anomalously and becomes peak-like, with extremely high absolute values of $\alpha$. This again indicates a change of the nature of the superconducting transition from second order to first order for magnetic field on the order of 4.6 T. A first order transition should result in a jump in the sample length, corresponding to a divergence of $\alpha$. Thus the peak-like signature in $\alpha$ indicates a broadened first order transition. Additional evidence for this interpretation comes from the isothermal magnetostriction experiments
displayed in the inset of Fig. 3. Whereas at $T=1.5\,\mbox{K}$ the kink in $\Delta l/l$ at $\rm H = 3.55\,\mbox{T}$ indicates a second-order phase transition, the jump in $\Delta l/l$, observed for $T = 0.2\,\mbox{K}$ at $\rm H = 4.86\,\mbox{T}$, provides clear evidence for the first-order nature of the transition.

How do our experimental results compare with theoretical predictions? Pauli paramagnetism leads to an upper limit for the magnetic field $H_p = \Delta_0/\sqrt2 \mu_B$, called Clogston paramagnetic limit~\cite{clogston:prl_62}, which superconductor can support. Here $\Delta_0$ is the superconducting energy gap, and $\mu_B$ is the Bohr magneton. Orbital effects of magnetic field also limit $H_{c2}$. The relative strength of the orbital pair breaking of magnetic field and Pauli limiting can be characterized by the parameter $\alpha = \sqrt 2 H_{c20}/H_p$, introduced by Maki~\cite{maki:pr_66}, where $H_{c20}$ is an orbital critical field in the absence of the Pauli limiting.  Maki's calculations~\cite{maki:pr_66} show that for $\alpha \ge 1$ the second order phase transition between the normal state and Abrikosov vortex state becomes unstable and changes to first order at a higher field, and if orbital effect is neglected ($\alpha = \infty$), this change takes place at the reduced temperature $t_0 = T_0/T_c = 0.55$.~\cite{maki:ptp_64} 

Also in the early 60's, an alternative theory, also based on Zeeman energy of electron spins in magnetic field, suggested that a new spatially inhomogeneous superconducting state, now called the Fulde-Ferrell-Larkin-Ovchinnikov (FFLO) state, may be stabilized close to the critical field $H_{c2}$ at which superconductivity is suppressed to zero~\cite{fulde_ferrell:pr_64,larkin_ovchinnikov:jetp_64}. FFLO state forms a wedge between the normal state and the homogeneous Abrikosov vortex state.  The tricritical point where the phase boundaries between the three states meet denotes the point of instability of second order phase transition at low field to the appearance of the FFLO wedge above the critical field within the FFLO scenario. Gruenberg and Gunther~\cite{gruenberg:prl_66} (GG) generalized the FFLO theory and included the orbital effect of magnetic field. The authors conclude that when the orbital pairbreaking effect is sufficiently small ($\alpha \ge 1.81$), 

\noindent
FFLO state can exist in a type-II superconductor. They also calculated the dependence of the reduced temperature $t_0$ of a tricritical point on the Maki parameter $\alpha$. For $\alpha = \infty$ (orbital pair breaking is ignored), $t_0 = 0.55$. Therefore, both Maki and FFLO scenarios give the same instability point for the second order phase transition in this limit. 

We do not see evidence for a second phase transition, which should be first order between FFLO state and a vortex state in FFLO scenario. However, we will make an assumption that we can use the result of the GG calculation of the instability point of the second order pase transition for arbitrary $\alpha$ within FFLO picture, to represent the instability of the second order phase transition within the Maki's scenario for arbitrary $\alpha$, since both theories give the same result in the limit $\alpha = \infty$, where calculations for Maki's scenario exist. To estimate $\alpha$ for \Co\, we use $\Delta_0 = 2.14 k_B T_c$  for a d-wave superconductor and obtain $H_p = 2.25 T_c\,{\rm T/K} = 5.2 \, \rm T$. The orbital critical field is $H_{c20} = 0.7 H_{c2}^\prime T_c = 13.2 \,\rm T$~\cite{movshovich:prl_01}. We therefore obtain $\alpha = 3.6$, corresponding to $t_0 = 0.35$ from Fig. 1, curve $b$ of Ref. \cite{gruenberg:prl_66}. Alternatively, it is possible to find $H_p$ from curve $a$ of Fig. 2 of Ref.~\onlinecite{gruenberg:prl_66}, which relates $H_{c2}(T=0)/H_p$ to $\alpha = \sqrt 2 H_{c20}/H_p$. Experimentally determined $H_{c2}= 4.95 \, \rm T$, and the only unknown parameter is $H_p$. We find $H_p = 5.8 \, \rm T$, $\alpha = 3.2$ and $t_0 = 0.33$. Both values of $t_0$ are very close to the value of $t_0 = 0.31$ observed experimentally. This agreement leads us to conclude that the first order phase transition is indeed due to the Pauli paramagnetism, confirming Maki's prediction of almost 40 years ago~\cite{maki:ptp_64,maki:pr_66}. 

The FFLO state has also attracted great attention, but its unambiguous observation has not been made. In the last decade the FFLO state was suggested to exist in heavy fermion UPd$_2$Al$_3$ (Ref.~\cite{gloos:prl_93}) and CeRu$_2$ (Ref.~\cite{huxley:jpcm_93}), based on thermal expansion and magnetization data, respectively. Subsequent research identified the magnetization feature in CeRu$_2$ as due to the flux motion~\cite{tenya:PhysicaB_99}, and the region of the suggested FFLO state in UPd$_2$Al$_3$ was shown to be inconsistent with theoretical model~\cite{norman:prl_93}. Most notably, no indication of the FFLO state was {\it ever} observed via specific heat measurement, which is the primary tool for identification of the thermodynamic details of the phase transition. \Co, according to theory, has all the prerequisite properties for the observation of the FFLO state, including that superconductor must be in clean limit. \Co\ satisfies this requirement as well, with quasiparticle mean free path $l_{tr} \approx 14 \xi_0$ at $T_c$~\cite{movshovich:prl_01}. Within superconducting state thermal conductivity divided by temperature grows by an order of magnitude as temperature is lowered to $T = 0.2 T_c$, and even at 30 mK (1\% of $T_c$) \Co\ is outside of the impurity band~\cite{movshovich:prl_01}. The  agreement between theoretical prediction and experimental observation of the critical point at $t_0$ shows that the pertinent physics of electronic spin Zeeman energy and Pauli effect drives the behavior of the system. However, within FFLO picture two transitions are expected: from normal state into FFLO state, likely to be second order, and from FFLO state into a usual Abrikosov vortex state via first order phase transition. We only observe one phase transition in \Co\ with field $H \parallel$ [001]. Therefore, FFLO state is not present in \Co, suggesting that FFLO state, if it exists, is much less stable than previously thought.

In summary, the superconducting transition in \Co\ with field $H \parallel [001]$ becomes first order below $T_0 = 0.31 T_c$. This is consistent with long standing theoretical predictions which take into account both orbital and spin interactions of superconducting electrons with magnetic field. We do not observe inhomogeneous superconducting FFLO state proposed theoretically almost 40 years ago~\cite{fulde_ferrell:pr_64,larkin_ovchinnikov:jetp_64,gruenberg:prl_66}. Instead, first order superconducting transition in \Co\ takes place from normal metal into mixed vortex state, in accord with scenario suggested by Maki~\cite{maki:ptp_64,maki:pr_66}.

We thank I. Vekhter, L. Boulaevskii, and K. Maki for stimulating discussions. Work at Los Alamos National Laboratory was performed under the auspices of the U.S. Department of Energy.


\end{document}